\documentclass[twocolumn,twoside,slac_two]{revtex4}
\usepackage{graphicx}
\usepackage{fancyhdr}
\newcommand{\simgt}%
{\,\hbox{\lower0.6ex\hbox{$\sim$}\llap{\raise0.6ex\hbox{$>$}}}\,}
\newcommand{\simlt}%
{\,\hbox{\lower0.6ex\hbox{$\sim$}\llap{\raise0.6ex\hbox{$<$}}}\,}
\newcommand{\invab}%
{{\rm ab}^{-1}}

\begin{document}

%Title of paper
\title{Future $e^+\,e^-$ flavour factories: detector challenges and physics expectations}

% Repeat the \author .. \affiliation  etc. as needed
%
% \affiliation command applies to all authors since the last
% \affiliation command. The \affiliation command should follow the
% other information

\author{Y. Ushiroda}
\affiliation{High Energy Accelerator Research Organization (KEK), Tsukuba, Japan}

\begin{abstract}
 Prospects of the two future $e^+\,e^-$ flavour factories are discussed.
 The detector designs and the technical challenges are described together with the
 motivating physics background.
\end{abstract}

%\maketitle must follow title, authors, abstract
\maketitle

\thispagestyle{fancy}

% body of paper here - Use proper section commands
% References should be done using the \cite, \ref, and \label commands
% Put \label in argument of \section for cross-referencing
%\section{\label{}}

\section{Introduction}
Since the beginning of the KEKB/Belle and the PEP-II/BaBar experiments, both $B$-factories
have been exploring the luminosity frontier. Both exceeded their design
luminosity ($1\times 10^{34}/cm^2/s$ for KEKB and $3\times 10^{33}/cm^2/s$
for PEP-II) and continued a steady operation.
Two factories integrated more than $1.3\,\invab$ of data in total, and
have reported a lot of important physics results every year. 
Some of the measurements reached already precision measurements,
however, there are quite a few of unsettled issues. Just as an example,
we have measured $\sin 2\phi_1$ with a precision of 4\,\% in $B\to J/\psi
K^0$ and related decay modes. On the other hand, we found that the effective
$\sin 2\phi_1$ measured in the penguin ($b\to sq\bar{q}$) decays such as
$B\to\phi K^0$ is possibly slightly smaller than $\sin 2\phi_1$. This
possible shift could be explained by a contribution from the new physics (NP)
beyond the standard model (SM); to confirm the shift with confidence, much more
statistics would be needed.
We wish to continue finding answers to those unsettled issues with
more data. To accumulate more data in a reasonable time-line, we need to
upgrade the accelerator and the detector.

%The LHC experiments are starting soon. The energy frontier experiments
%would discover the NP particles. In that case, we need to study the flavour
%parameters of the NP, which is possible with the upgraded $B$-factory, but
%not in the energy frontier experiments. If, unfortunately, the energy
%frontier experiments does not find the NP, the alternative way to keep
%searching for the NP will be the upgraded $B$-factory.

There are three facilities that will study $B$ physics including future
projects: the LHC$b$~\cite{LHCb}, the
KEKB upgrade (Super KEKB~\cite{SuperKEKB}), and the
Super$B$~\cite{SuperB} in Italy.
The LHC$b$ will produce lots of
beauty particles including $B_s$ and $b$-baryons in a hadronic manner.
It is good for accumulating data, but the background environment does
not allow us to study the modes with neutral
particles in the final state. One of the important roles of the Super KEKB
and the Super$B$ are
to do analyses with neutral particles in the final state in a clean
environment of the $e^+\,e^-$ collider.
The complemental roles of the LHC$b$ and the other two is 
well summarized in the preprint by I.~Bigi~\cite{Bigi:2008qr}.
The target of this presentation is to describe
the common things and the differences of the Super KEKB and the Super$B$.

\section{Accelerator and Beam Background}
%Before discussing on the detector and the physics, we must understand
%the accelerator in order to know how much and by when we can accumulate
%our signals, and also to know how the beam background will be.
%In fact,
The approaches to achieve high luminosity are very different in
the Super KEKB and the Super$B$.
In the Super KEKB, luminosity gain is obtained mainly from the higher
beam current. On the other hand, the Super$B$ aims for a high luminosity
by squeezing the beam size drastically, while the beam currents are kept
at the same level as in the current $B$-factories. Though both approaches
are challenging in the different viewpoints, Super KEKB for the high
current operation, and the Super$B$ for the nano beam operation, 
it is often said that Super KEKB is a brute force but
a steady and adiabatic method, and the Super$B$ is a brand-new and more
sophisticated method.

The target luminosity of the Super KEKB is
$8\times 10^{35}\,cm^{-2}s^{-1}$, and the integrated luminosity for the
discussion of physics is $10\,\invab$ as the first target, and then
we aim for $50\,\invab$ .
On the other hand, the luminosity of Super$B$ is expected to begin with
$1\times 10^{36}\,cm^{-2}s^{-1}$ and it can be improved by a few times.
The integrated luminosity will reach $80\,\invab$ in seven years.
However, one should be aware that the lattice design and the beam-beam
simulation that support the design luminosity is not as matured as in
the case of the Super KEKB, which should be improved in the coming years.

In the Super KEKB scheme, the most severe beam-induced background is from
the beam-gas scattering. Due to $N$ times higher beam current, the vacuum around
the interaction region becomes $N$ times worse, provided the
evacuation power is the same; hence, the amount of the beam-gas background will
be $N^2$ times larger than the present level. In total, at the full spec of the
machine, we will receive some 20 times more background than the present level.
Although there are several ideas to improve the background, the detector
is designed so that it works under such high background conditions.

In the case of the Super$B$, one of the biggest worry is in the Touschek
background, which is an intra-beam scattering enhanced in the narrow
beams of the Super$B$.
According to a simulation study, the Touschek background could be reduced by
three to four orders of magnitudes with a new lattice and collimator;
with which detector designs are feasible.

%In general, the amount of the beam background depends on the design of
%IR. As we finalize the machine, we should pay special attentions to
%the beam background, not only the luminosity.

\section{Detector Design and Expected Physics Reach}
As discussed in the previous section, in the case of the Super KEKB, the
detector (sBelle\footnote{The name of the detector is to be
determined. The 'sBelle' is not an authorized name, used only within this
proceedings for convenience.})
must work under typically 20 times higher background; in the
case of the Super$B$, the condition is supposed to be milder
because of the lower beam current.
What we do in the detector design against the high
background is simple in principle: to use faster technology and/or to
have smaller segmentation of the sensors. The idea is to avoid the overlap of
the particle hits in time and/or in space.

The baseline design of the sBelle sub-detectors is as follows.
We will have a 1.5\,cm radius beam pipe and 6 layers of silicon vertex
detectors (SVD).
We will adopt a new readout ASIC, APV25, which has
16 times shorter shaping time than the present readout chip (VA1TA).
For the central drift chamber (CDC), we
will have a small cell chamber with about 15 thousand sense wires. According to
the simulation studies, by improving the tracking algorithm for the CDC,
and also with the help of the SVD stand-alone tracking,
we can improve the performance of the charged track
reconstruction even under 20 times more background.
For the particle identification (PID) detector, we will not be able to use
the time-of-flight (TOF) counters which are based on the scintillation
counters, because of the high counting rate. The possible device should be
Cherenkov detectors since they are insensitive to high energy photons.
Among several options for the barrel PID detectors, the baseline is the
time-of-propagation (TOP) counter, which reconstruct the Cherenkov ring
image in one of the spatial coordinates and in time. The PID detector in the
forward end-cap is a
proximity focusing ring imaging Cherenkov counter with the aerogel
radiators (A-RICH). Thallium doped CsI crystals are used for the
electromagnetic calorimeter (ECL). The end-caps will be replaced with faster
crystals such as pure CsI. The waveforms of the signals are sampled and
fitted to resolve the overlap of multiple hits. According to a simulation
study, the waveform fitting has an effect to suppress the background
clusters by a factor of seven.
The $K_L$ and muon detector (KLM) is based on resistive-plate-counter in
the barrel and scintillator in the end-caps. More details of the
detector design are described elsewhere~\cite{bib:sBelleLoI}.

% figure ? 

The design of the Super$B$ detector is based on the BaBar detector.
The concept of the upgrade is similar to that for the sBelle, but with
somewhat milder condition of the beam background.
Aged components such as the silicon vertex tracker (SVT), the drift
chamber (DCH) are to be renewed. The DIRC counter can be reused as the
PID detector, possibly with new photon sensors that allow
a smaller stand-off box.
The forward end-cap of the
electromagnetic calorimeter (EMC) will be replaced with faster crystals
such as LYSO. The $K_L$ and muon detection will be based on
scintillators. More details of the detector design are described
elsewhere~\cite{Bona:2007qt}.

In the following subsections, some of the key features of the detector
upgrade are discussed. Possible physics reaches related to the upgrade
options are also described.

\subsection{Time-Dependent $CP$ Analyses and Vertex Detector}
\begin{figure}
\includegraphics[width=0.9\columnwidth]{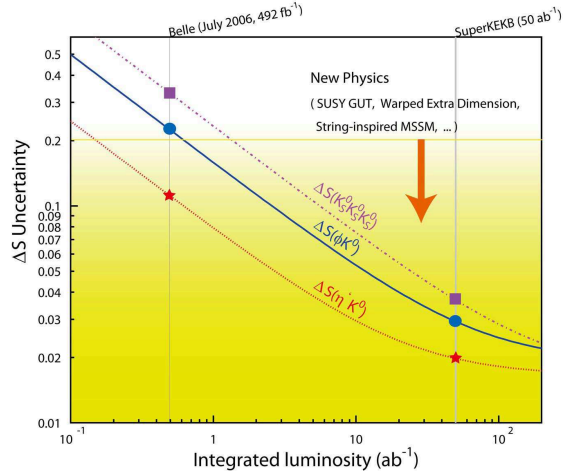}%
\caption{\label{fig:b2sqq} Expected sensitivity of the $\mathcal{S}_{\phi
 K^0}$, $\mathcal{S}_{\eta^\prime K^0}$ and $\mathcal{S}_{K^0K^0K^0}$
 with regard to the accumulated luminosity.
}
\end{figure}

%In the present $B$-factories, the vertex detector is extremely important
%to measure the time-dependent $CP$ asymmetry in $B\to J/\psi K_S$ to
%determine $\sin 2\phi_1$~\cite{bib:sin2phi1}. In the upgrade, the $\sin
%2\phi_1$ will have been measured already precisely, however, ........
In the upgrade, the vertex detector is important for the time-dependent
$CP$ (TCP) asymmetry measurements.
Especially, the TCP in $b\to s\bar{q}q$ penguin decays should be measured
precisely, because it could be affected by the NP.
Figure~\ref{fig:b2sqq} shows the expected sensitivities of the
TCP parameters ($\mathcal{S}$) in $B^0\to\phi K^0$,
$B^0\to\eta^\prime K^0$, and $B^0\to K^0 K^0 K^0$ decays; the
extrapolation is based on the analyses of Belle data~\cite{Chen:2006nk}.
With $50\,\invab$, the uncertainty of the measured
$\mathcal{S}$ reaches the level of the current theoretical uncertainty
(0.03). It also indicates that the measurement will be limited by the
systematic uncertainty after $50\,\invab$.
To go beyond $50\,\invab$,
we must seriously consider how to reduce the intrinsic systematics from
the sources such as the alignment of the vertex detector.

The key issue to improve these analyses is the proper time ($\Delta t$)
resolution, which is determined from the vertex resolution ($\Delta z$) and the boost
factor ($\beta\gamma$) of the $\Upsilon(4S)$ rest frame in the
laboratory frame as $\Delta t = \Delta z/\beta\gamma c$.
We can achieve a better
vertex resolution by having the first sensor of the vertex detector as
close to the interaction point (IP) as possible.
In the Super$B$, since the boost factor is small, the desire to put the
first sensor closer is ardent. A beam pipe of 1.0\,cm radius that allows
to place the innermost layer at the radius of 1.2\,cm is proposed.
The technical challenges for this are (1) higher background at the
sensor, (2) the stronger synchrotron radiation fan,
and (3) the higher-order-mode heating of the beams.

The TCP in $B^0\to K^0_S\pi^0\gamma$ is also an interesting mode to search for
the NP effects from a different aspect~\cite{Atwood:1997zr}.
In this decay mode, since there is no charged particle in the final
state of the signal $B^0$ decays that comes from the IP, 
we need to determine the $B^0$ decay position from the $K^0_S$
trajectory which is determined from the daughter $\pi^+$ and $\pi^-$
trajectories.
Therefore, a larger vertex detector that allows more
$K^0_S\to\pi^+\pi^-$ decays inside the vertex detector volume is
preferred.

From the technical point of view, if we build a larger vertex detector in
a conventional ladder structure, the detector capacitance becomes larger
($\simgt 60\,pF$), which means a higher noise in the readout chip.
Since a faster readout chip, which we need so that the background
overlap can be avoided, is in general more susceptible to higher input
capacitance, here is a technical difficulty. One way to avoid this
dilemma is to abandon the conventional ladder structure and to put the
readout chips on the sensor itself to reduce the capacitance.
Special care should be taken for the material inside the sensitive
area, and also for the cooling and the influence of the heat-cycling on
the sensor.

\begin{figure}
\includegraphics[width=0.9\columnwidth]{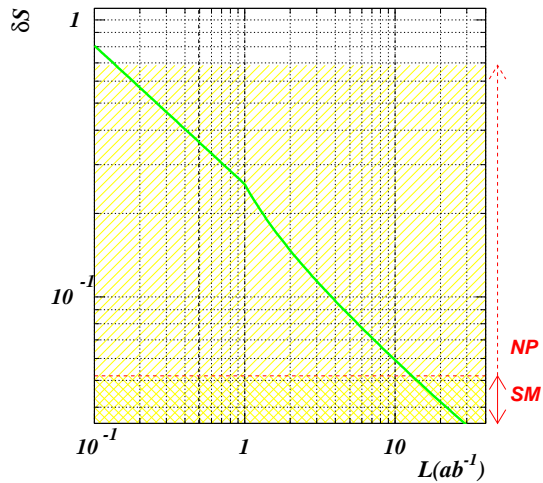}%
\caption{\label{fig:b2sgam} Expected sensitivity of the $\mathcal{S}_{K^{\ast0}\gamma}$
 with regard to the accumulated luminosity.
}
\end{figure}

Figure~\ref{fig:b2sgam} shows the expected sensitivity of the TCP
parameter ($\mathcal{S}$) in the $B^0\to K^\ast(K^0_S\pi^0)\gamma$ decay
based on the analysis of Belle data~\cite{Ushiroda:2006fi}.
Thanks to the detector upgrade that appears as a kink at
$1\,\invab$, we can reach sensitivities close to the level of the SM
theoretical uncertainty with some $10\,\invab$.

\subsection{Particle Identification}
The PID of the Belle detector has been working without any serious
troubles from the very beginning until now. The performance is good
enough so that we could pursue many analyses by discriminating kaons from pions
and vice versa. One pity is that the kaon fake rate in the pion
candidates is higher than the case of the BaBar. This makes the analyses
that require kaon veto more difficult; $B^0\to\rho^0(\to\pi^+\pi^-)\gamma$ is one of
such analyses where the $B^0\to K^{\ast0}(\to K^+\pi^-)\gamma$ decay with the
$K^+$ misidentified as the $\pi^+$ contributes as a severe background.
To improve the performance, we will completely renew the PID detectors
both in the barrel and in the forward end-cap. Two additional reasons to
replace the PID detectors are: because the TOF will not work under high
background condition of the Super KEKB, and because the ACC container and
the photomultiplier are massive which is bad for the photon reconstruction in
the calorimeter.

The baseline option of the barrel PID is the TOP counter, which measures
one spatial coordinate with a few mm precision and the time of
Cherenkov light arrival with the time resolution of $\simlt 40\,ps$.
In addition to the excellent time resolution, the photo-sensor is
required to work under 1.5\,T magnetic field. A micro-channel-plate
PMT (MCP-PMT) is a promising candidate for such a purpose.

The baseline option of the end-cap PID is the A-RICH. The three
candidate photo-sensors, the hybrid avalanche photo-diode, the MCP-PMT,
and the multi-pixel photon counter (MPPC) have been extensively tested at the
test-beam line at KEK. In the test, clear ring images are successfully
observed.
In the case we use the MCP-PMT, additional discrimination power is
obtained because the time resolution is good enough to measure the TOF
difference of kaons and pions.

\begin{figure}
\includegraphics[width=0.9\columnwidth]{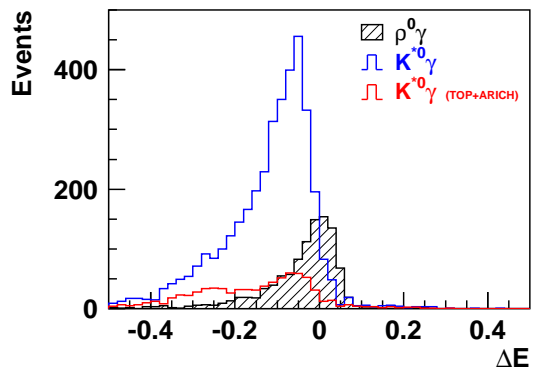}%
\caption{\label{fig:b2rhogam}
The $\Delta E$ distributions for the $B^0\to\rho^0\gamma$ signal
 (hatched) and the $K^{\ast0}\gamma$ background. The blue(red) line
 shows before(after) the upgrade of the PID.
}
\end{figure}

Figure~\ref{fig:b2rhogam} shows the $\Delta E$ distributions for the
$\rho^0\gamma$ signal and the $K^{\ast0}\gamma$ background in the
$B^0\to\rho^0\gamma$ analysis. By the upgrade of the PID detector, we
can significantly reduce the $K^{\ast0}\gamma$ background.
The sensitivity gain with this improvement is equivalent to 83\,\% more
luminosity.

% SuperB PID

\subsection{Neutrinos and Hermeticity}

\begin{figure}
\includegraphics[width=0.8\columnwidth]{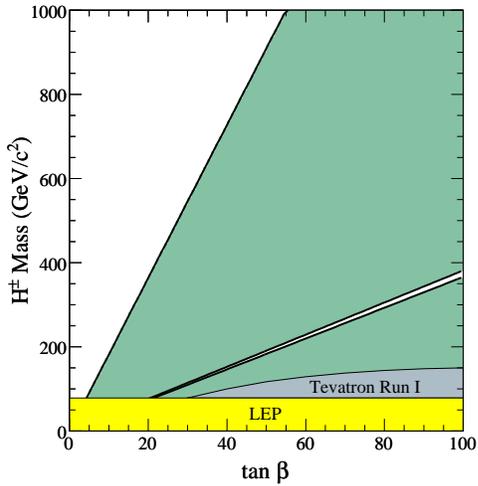}%
\caption{\label{fig:tanb_mH_exclude_50}
Exclusion region in the $\tan\beta$ vs. $H^\pm$ mass plane with
 $50\,\invab$ of data. Pale sea-green region would be excluded by $B\to\tau\nu$.
}
\end{figure}

\begin{figure}
\includegraphics[width=0.8\columnwidth]{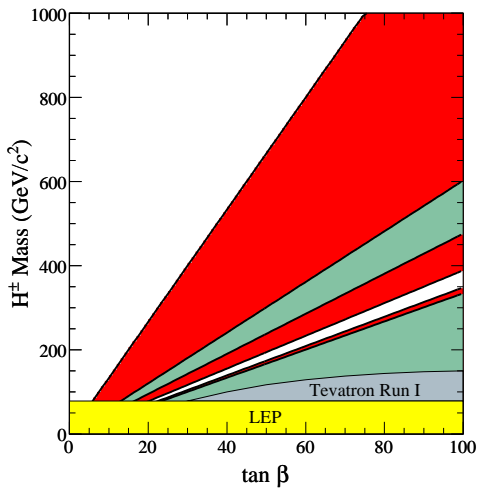}%
\caption{\label{fig:tanb_mH_discovery_50}
 Discovery region in the $\tan\beta$ vs. $H^\pm$ mass plane with
 $50\,\invab$ of data. Red region is where we could discover a
 charged Higgs with $5\,\sigma$, pale sea-green region has already been
 excluded.
}
\end{figure}

Another important type of measurements in the future $e^+\,e^-$
$B$-factories are the studies of the decay channels that contain
neutrinos in the final state.
Figure~\ref{fig:tanb_mH_exclude_50} shows the possible exclusion region
in the $\tan\beta$ vs. $H^\pm$ mass plane with $50\,\invab$ of data,
where $B^0\to\tau\nu$ provides a very powerful constraint as shown in
the pale sea-green region.
It could result in a $5\,\sigma$ discovery of the charged Higgs for some
cases as shown in the red region in Fig.~\ref{fig:tanb_mH_discovery_50}.

In this type of analyses, one of the most important issues for the
detector is the hermeticity. If the daughter particles are 
not detected because they pass through the insensitive region of the
spectrometer, 
we can not distinguish those missing particles and
the signal neutrinos; thus we suffer from a peaking type of background.
One idea to improve the detector hermeticity is to place detectors around
the forward and the backward beam holes.
From the simulation study of the $B\to K^{(\ast)}\nu\nu$ decays,
by reconstructing tracks by three layers of the pixel disks that
cover the beam hole,
we could eliminate 68.5\,\% of the peaking background while sacrificing
only 0.3\,\% of the signal. The technical challenges for this are:
(1) high background as it is close to the beam, (2) very limited
space for the detector and the cables, and (3) to understand the
magnetic field well despite it is close to the final focusing magnets.

\section{Conclusion}
Two future $e^+\,e^-$ flavour factories, the Super KEKB and
the Super$B$, aim to improve the measurements that have been performed by
the Belle and the BaBar collaborations.
The Super KEKB will increase the beam currents to increase the luminosity,
while the Super$B$ will squeeze the beam size to increase the luminosity
even higher than the Super KEKB.
The detectors must handle higher background than the detectors
of the present $B$-factories;
in the case of Super KEKB, 20 times higher background is
expected at the final design luminosity of the machine. A careful design of the
detector and the machine is essential for both cases. Further
improvements in the detector performance such as the vertex resolution,
the $K^0_S$ reconstruction efficiency with the vertex information,
the PID, the hermeticity are foreseen, although each of the improvements
is technically challenging. Two projects aim for the higher luminosity
in very different approaches, which is good as a strategy to diversify
the risk.

% If you have acknowledgments, this puts in the proper section head.
%\bigskip % extra skip inserted
\begin{acknowledgments}
 I am grateful to F. Forti for the information on the Super$B$.
 The slides and the comments from him were very
 helpful to compose this presentation.
 For the Super KEKB detector, I would like to thank the members of the
 sBelle design task force for the extensive simulation studies.
\end{acknowledgments}

\bigskip % extra skip inserted
% Create the reference section using BibTeX:
%\bibliography{basename of .bib file}
%\begin{thebibliography}{9}   % Use for  1-9  references

\end{document}